\documentclass[prd,aps,showpacs,tightenlines,nofootinbib,preprint]
{revtex4}
\newcommand{\bear}{\begin{array}}  \newcommand{\eear}{\end{array}}
\newcommand{\bea}{\begin{eqnarray}}  \newcommand{\eea}{\end{eqnarray}}
\newcommand{\beq}{\begin{equation}}  \newcommand{\eeq}{\end{equation}}
\newcommand{\bef}{\begin{figure}}  \newcommand{\eef}{\end{figure}}
\newcommand{\bec}{\begin{center}}  \newcommand{\eec}{\end{center}}

\newcommand{\bib}{\bibitem}

\def\APJ#1#2#3{Astrophys. J. {\bf #1}, #2 (19#3)}
\def\APJJ#1#2#3{Astrophys. J. {\bf #1}, #2 (20#3)}

\def\ARAA#1#2#3{Ann. Rev. Astron. Astrophys. {\bf#1}, #2 (19#3)}
\def\ARAAA#1#2#3{Ann. Rev. Astron. Astrophys. {\bf#1}, #2 (20#3)}

\def\IJMPDD#1#2#3{Int. J. Mod. Phys. D {\bf #1}, #2 (20#3)}

\def\JHEPP#1#2#3{J. High Energy Phys. {\bf #1}, #2 (20#3)}

\def\MNRASS#1#2#3{Mon. Not. R. Astron. Soc. {\bf #1}, #2 (20#3)}

\def\NPB#1#2#3{Nucl. Phys. {\bf B#1}, #2 (19#3)}

\def\PLB#1#2#3{Phys. Lett. B {\bf #1}, #2 (19#3)}
\def\PLBB#1#2#3{Phys. Lett. B {\bf #1}, #2 (20#3)}
\def\PL#1#2#3{Phys. Lett. {\bf #1}, #2 (19#3)}

\def\PR#1#2#3{Phys. Rev. {\bf #1}, #2 (19#3)}

\def\PRD#1#2#3{Phys. Rev. D {\bf #1}, #2 (19#3)}
\def\PRDD#1#2#3{Phys. Rev. D {\bf #1}, #2 (20#3)}

\def\PRL#1#2#3{Phys. Rev. Lett. {\bf#1}, #2 (19#3)}
\def\PRLL#1#2#3{Phys. Rev. Lett. {\bf#1}, #2 (20#3)}
\def\PRS#1#2#3{Proc. Roy. Soc. {\bf A#1}, #2 (19#3)}

\def\PRTT#1#2#3{Phys. Rep. {\bf#1}, #2 (20#3)}

\def\RPP#1#2#3{Rep. Prog. Phys. {\bf #1}, #2 (19#3)}

\def\RMPP#1#2#3{Rev. Mod. Phys. {\bf #1}, #2 (20#3)}

\def\Vec#1{\mbox{\boldmath $#1$}}

\begin{document}

\title{
Property of the spectrum of large-scale magnetic fields from inflation 
      }

\author{Kazuharu Bamba\footnote{
Present address:    Department of Physics, Kinki University, 
                    Higashi-Osaka, Osaka 577-8502, Japan  \\
Electronic address: bamba@phys.kindai.ac.jp
}
%Research Fellow of the Japan Society for the Promotion of Science}
%\footnote{E-mail: bamba@yukawa.kyoto-u.ac.jp}
}  
\affiliation{
Yukawa Institute for Theoretical Physics, 
Kyoto University, 
Kyoto 606-8502, 
Japan}

%\date{\today}

%%%%%%%%%%%%%%%%%%%%%
%  Abstract
%%%%%%%%%%%%%%%%%%%%%
\begin{abstract}

The property of the spectrum of large-scale magnetic fields generated due to 
the breaking of the conformal 
invariance of the Maxwell theory through some mechanism 
in inflationary cosmology is studied. 
It is shown that the spectrum of the generated magnetic fields should 
not be perfectly scale-invariant but be slightly red 
so that the amplitude of large-scale magnetic fields 
can be stronger than $\sim 10^{-12}$G at the present time.  
%%%%%
This analysis is performed by assuming the absence of 
amplification due to the late-time action of some dynamo 
(or similar) mechanism. 
%%%%%

\end{abstract}

%%%%%%%%%%%%%%%%%%%%%

%----------------------------
\pacs{98.80.Cq, 98.62.En 
%Keywords: 
%magnetic fields, 
%inflation, 
%physics of the early universe
}
\hspace{13.0cm} YITP-07-13
%----------------------------

\maketitle

%%%%%%%%%%%%%%%%%%%%%%%%%%%
%%%  Sec. I 
%%%%%%%%%%%%%%%%%%%%%%%%%%%
\section{Introduction}

It is observationally known that there exist magnetic fields with the field 
strength $\sim 10^{-6}$G on $1-10$kpc scale in galaxies of all types 
(for detailed reviews, 
see 
\cite{Sofue, Kronberg1, Grasso, Carilli1, Widrow, Giovannini1, Semikoz1}) 
and in galaxies at cosmological distances \cite{Kronberg2}. 
Furthermore, 
magnetic fields in clusters of galaxies with the field strength 
$10^{-7}-10^{-6}$G on 10kpc$-$1Mpc scale 
have been observed \cite{Kim1}.   
It is very interesting and mysterious that magnetic fields in clusters of 
galaxies are as strong as galactic ones and that the coherence scale may be 
as large as $\sim$Mpc.  The origin of these magnetic fields is not well 
understood yet.  
Although galactic dynamo mechanisms \cite{EParker} have been proposed 
to amplify very weak seed magnetic fields up to $\sim 10^{-6}$G, 
they require initial seed magnetic fields to feed on.  
Moreover, the effectiveness of the dynamo amplification mechanism 
in galaxies at high redshifts or 
clusters of galaxies is not well established.  

Proposed generation mechanisms of seed magnetic fields 
fall into two broad categories.  One is astrophysical processes, 
%e.g., plasma instabilities, such as the Weibel instability \cite{PI}, 
e.g., the Biermann battery mechanism \cite{Biermann1} 
and the Weibel instability \cite{PI}, which is a kind of 
plasma instabilities, and 
the other is cosmological processes in the early Universe, 
e.g., the first-order cosmological electroweak phase 
transition (EWPT) \cite{Baym}\footnote{In Ref.~\cite{Durrer}, 
it has been pointed out 
that causally produced stochastic magnetic fields on large scales, e.g., 
during EWPT or even later, are much stronger suppressed than usually assumed.
}, quark-hadron phase transition (QCDPT) 
\cite{Quashnock} (see also \cite{Boyanovsky1}), 
and the generation of the magnetic fields from primordial 
density perturbations before the epoch of recombination 
\cite{Notari1, Ichiki1, Gopal1, Siegel1, Berezhiani1, Kobayashi07}.  
However, 
it is difficult for these processes to generate 
the magnetic fields on megaparsec scales with the sufficient field strength 
to account for the observed magnetic fields in galaxies and clusters of 
galaxies without requiring any dynamo amplification.  

The most natural origin of such a large-scale magnetic field is 
electromagnetic quantum fluctuations generated in the inflationary stage 
\cite{Turner}. 
%(for a review of inflation, see Refs.~\cite{Linde1, Kolb}) 
This is because inflation naturally produces effects on very large scales, 
larger than Hubble horizon, starting from microphysical processes 
operating on a causally connected volume.  
Since the Friedmann-Robertson-Walker (FRW) metric 
usually considered is conformally flat and 
the classical electrodynamics is conformally invariant, 
the conformal invariance of the Maxwell theory must have been 
broken in the inflationary stage\footnote{
In Ref.~\cite{Maroto1}, the breaking of conformal flatness of the 
FRW metric induced by the evolution of scalar metric perturbations at 
the end of inflation has been discussed.} 
in order that electromagnetic quantum fluctuations could be generated 
at that time \cite{Parker}.  
Several breaking mechanisms therefore have been proposed 
\cite{Turner, RF^2, Ratra, Scalar, Charged-Scalar, 
ScalarED, Amplification, 
Dolgov1, Bertolami1, Gasperini1, Prokopec1, Enqvist1, 
Bertolami2, Ashoorioon1, Bamba1, Bamba2, Bamba3}.  

In the present paper we discuss the spectrum of large-scale magnetic fields 
generated due to the breaking of the conformal invariance of the Maxwell 
theory through some mechanism in inflationary cosmology.  
%%%%%
In particular, we perform the analysis of the spectrum of 
large-scale magnetic fields 
by assuming the absence of amplification due to the late-time action 
of some dynamo (or similar) mechanism. 
%%%%%
We use units in which $k_\mathrm{B} = c = \hbar = 1$.\ 

This paper is organized as follows.  
In Sec.\ II we consider the constraint on the amplitude of 
large-scale magnetic fields with a scale-invariant spectrum.  
In Sec.\ III we discuss the spectrum of large-scale 
magnetic fields. 
Finally, Sec.\ IV is devoted to a conclusion.

%%%%%%%%%%%%%%%%%%%%%%%%%%%
%%%  Sec. II 
%%%%%%%%%%%%%%%%%%%%%%%%%%%
\section{Constraint on the amplitude of 
large-scale magnetic fields with a scale-invariant spectrum}

We assume the spatially flat Friedmann-Robertson-Walker (FRW) space-time with 
the metric 
\begin{eqnarray}
 {ds}^2 = g_{\mu\nu}dx^{\mu}dx^{\nu} =  -{dt}^2 + a^2(t)d{\Vec{x}}^2,
\label{eq:1}
\end{eqnarray} 
where $a(t)$ is the scale factor.  
The background Friedmann equation is given by 
\begin{eqnarray}
  H^2 &=&  \left( \frac{\dot{a}}{a} \right)^2 = 
\frac{8\pi}{3{M_{\mathrm{Pl}}}^2} {\rho}_{\phi}
\equiv {H_{\mathrm{inf}}}^2,  
\label{eq:2} \\[1.5mm]
 {\rho}_{\phi} &=& \frac{1}{2}{\dot{\phi}}^2 + U[\phi], 
\label{eq:3}  
\end{eqnarray} 
where a dot denotes a time derivative.  Here, $H$ is the Hubble parameter, 
${\rho}_{\phi}$ is the energy density of the inflaton field, $U[\phi]$ is 
the inflaton potential, and 
$M_{\mathrm{Pl}} = G^{-1/2} = 1.2 \times 10^{19}$GeV is the Planck mass.  
Moreover, $H_{\mathrm{inf}}$ is the Hubble constant in the inflationary 
stage.  

It is well known that for a minimally coupled scalar field in de Sitter space, 
there are fluctuations in that field with energy density corresponding to that 
of a thermal bath at the Gibbons-Hawking temperature, 
$H_\mathrm{inf}/(2\pi)$ 
\cite{Gibbons,Bunch}.  
According to Turner and Widrow \cite{Turner}, it is reasonable to assume that 
all quantum fields in de Sitter space, in particular the electromagnetic field 
(if the conformal invariance of the Maxwell theory is broken through 
some mechanism in the inflationary stage), are excited with an energy density 
of order $\left[ H_\mathrm{inf}/(2\pi) \right]^4$.  

Here we consider the case in which the conformal invariance of the Maxwell 
theory is broken through some mechanism in the inflationary stage and 
then magnetic fields whose origin is electromagnetic quantum fluctuations 
amplified during inflation are generated.  
Moreover, we assume that the spectrum of the generated magnetic fields is 
scale-invariant.  
%%%%%
The quantum degree of freedom is the photon field which will lead to the same 
amount of electric and magnetic energy.  
After inflation, however, when the conductivity of the Universe 
becomes high, the electric fields will be dissipated 
and only the magnetic fields survive.  
%%%%%
It is conjectured from the above discussion that the energy density of the 
magnetic fields, $\rho_B$, is 
$\rho_B \lesssim \left[ H_\mathrm{inf}/(2\pi) \right]^4$.  
From this relation and Eq.\ (\ref{eq:2}), we find 
\begin{eqnarray}
 \frac{\rho_B}{\rho_\phi} \lesssim \frac{1}{6{\pi}^3} 
\left( \frac{H_{\mathrm{inf}}}{M_{\mathrm{Pl}}}
\right)^2.
\label{eq:4}
\end{eqnarray} 

The upper limit on $H_{\mathrm{inf}}$ is determined by the 
observation of the anisotropy of the 
cosmic microwave background (CMB) radiation.  
Using the Wilkinson Microwave Anisotropy Probe (WMAP) three year data on temperature fluctuation \cite{Spergel06}, 
%, which is consistent with the Cosmic Background Explorer (COBE) data, 
we can obtain a constraint on $H_{\mathrm{inf}}$ from 
tensor perturbations \cite{Rubakov,Abbott}, 
\begin{eqnarray}
\frac{H_{\mathrm{inf}}}{M_{\mathrm{Pl}}} \leq 4.9 \times 10^{-5}.
\label{eq:5} 
\end{eqnarray} 

Here we consider the case in which after inflation 
the Universe is reheated immediately at $t=t_\mathrm{R}$ 
% (see e.g.\ \cite{KLS} for an efficient mechanism of reheating) 
and then all the energy of the inflaton is reduced to radiation.  
Moreover, the conductivity of the Universe $\sigma$ 
is negligibly small during inflation, because there are few charged particles 
at that time. 
After reheating, however, a number of charged particles are produced, so that 
the conductivity immediately jumps to a large value:\ 
$\sigma \gg H \hspace{1.5mm} 
(\hspace{0.5mm}t \geq t_\mathrm{R}\hspace{0.5mm})$ 
and hence it is always much larger 
than the Hubble parameter at that time in the radiation-dominated stage and 
the subsequent matter-dominated stage.  
This assumption is justified by a microphysical analysis \cite{Turner}.  
%%%%%
Consequently, as stated above, 
for a large enough conductivity at the instantaneous reheating stage, 
the electric fields accelerate charged particles and dissipate.  
%%%%%
Furthermore, we consider the case in which after reheating $\rho_B$ is not 
supplied with any energy by some fields through the coupling between those 
fields and electromagnetic fields 
and hence $\rho_B$ evolves in proportion to $a^{-4}(t)$.  
In this case, after reheating the ratio of the energy density of the magnetic 
fields to that of the radiation remains constant.  
Consequently, 
it follows from Eqs.\ (\ref{eq:4}) and (\ref{eq:5}) 
that the energy density of the magnetic fields at the present time $t_0$ is 
\begin{eqnarray} 
 \rho_B(t_0) \lesssim \frac{ \left(4.9 \times 10^{-5} \right)^2 }
{6{\pi}^3} {\rho}_{\gamma 0}, 
\label{eq:6} 
\end{eqnarray} 
where ${\rho}_{\gamma 0}$ is the present energy density of the CMB radiation.  
Using Eq.\ (\ref{eq:6}), 
${\rho}_{\gamma 0} = 2.1 \times 10^{-51} (T_{\gamma 0}/2.75)^4 \hspace{0.5mm}
{\mathrm{GeV}}^4$ \cite{Kolb}, where 
$T_{\gamma 0} \approx 2.73 \hspace{0.5mm} \mathrm{K}$ 
is the present temperature of the CMB radiation, 
and $1 [\mathrm{G}]^2/(8 \pi) = 1.9 \times 10^{-40} {\mathrm{GeV}}^4$, 
we find that 
the present magnetic field, $B(t_0)$, is 
$B(t_0) \lesssim 3.3 \times 10^{-12}$G. 

It is known that the required strength of the cosmic magnetic fields at the 
structure formation, adiabatically rescaled to the present time, is 
$10^{-10}-10^{-9} \mathrm{G}$ in order to explain the observed magnetic fields 
in galaxies and clusters of galaxies without dynamo amplification mechanism.  
If the spectrum of the generated magnetic fields is perfectly 
scale-invariant, 
it follows from the above consideration that 
the present strength of the magnetic fields is at most or smaller than 
$\sim 10^{-12}$G.  
Hence the spectrum of the magnetic fields should not 
be perfectly scale-invariant but be tilted in order that 
the amplitude of the large-scale magnetic fields 
can be as large as $10^{-10}-10^{-9} \mathrm{G}$ at the present time, 
so that the observed magnetic fields could be explained through only 
adiabatic compression without requiring any dynamo amplification.

%%%%%%%%%%%%%%%%%%%%%%%%%%%
%%%  Sec. III 
%%%%%%%%%%%%%%%%%%%%%%%%%%%
\section{Spectrum of large-scale magnetic fields}

Next, we discuss the spectrum of large-scale magnetic fields.  
We consider the case in which the Fourier components of 
the root-mean-square (rms) of the magnetic fields at the present time, 
$\left| B(k, t_0) \right|$, are given by 
\begin{eqnarray}
\left| B(k, t_0) \right|^2  \equiv  A \left( \frac{k}{k_\mathrm{c}} \right)^n. 
\label{eq:7} 
\end{eqnarray}
Here, $k = 2\pi/L$ is comoving wave number and 
$L$ is the comoving scale of the magnetic fields.  
Moreover, $n$ is the spectral index of $\left| B(k, t_0) \right|^2$ and 
$A$ is a constant.  
Furthermore, 
$k_\mathrm{c} = 2\pi/L_\mathrm{c}$, where 
$L_\mathrm{c}$ is the scale on which the field strength of 
the magnetic fields is equal to $10^{-12}$G. 
Multiplying $\left| B(k, t_0) \right|^2$ 
by phase-space density:\ $4\pi k^3/(2\pi)^3$, 
we obtain the magnetic fields 
in the position space at the present time 
\begin{eqnarray}
\left| B(L, t_0) \right|^2  =  \frac{Ak_\mathrm{c}^3}{2\pi^2} 
 \left( \frac{L_\mathrm{c}}{L} \right)^{n+3}. 
\label{eq:8} 
\end{eqnarray}
Using Eq.\ (\ref{eq:8}) and taking into account the fact that 
$L_\mathrm{c}$ is the scale on which the field strength of the 
magnetic fields is equal to $10^{-12}$G, we find 
\begin{eqnarray}
\left| B(L_\mathrm{c}, t_0) \right|^2  =  \frac{Ak_c^3}{2\pi^2} = 
\left( 10^{-12} \hspace{1mm} \mathrm{G} \right)^2. 
\label{eq:9} 
\end{eqnarray}

Here we consider the case in which the spectrum of the magnetic fields 
is red, i.e., $n < -3$.  
%%%%%
We here note the following point: 
In the case of a red spectrum, a large scale cutoff of the magnetic fields 
is needed, otherwise the total energy of the magnetic fields is infinite.  
Although a mechanism which leads to this infrared cutoff 
(which is determined by inflationary physics) 
is necessary, we have not found it yet.  
In the following discussion, therefore, we assume that a infrared cutoff of 
the magnetic fields is realized by some mechanism.  
%%%%%
In this case, if the current amplitude of the magnetic fields on 1Mpc scale 
is larger than $10^{-10} \mathrm{G}$, 
$L_\mathrm{c} < 1 \mathrm{Mpc}$.  
It follows from Eq.\ (\ref{eq:8}) that 
in order that the current amplitude of the magnetic 
fields on 1Mpc scale can be stronger than $10^{-10} \mathrm{G}$, 
$\left| B(L= 1 \mathrm{Mpc}, t_0) \right|^2 >  
\left( 10^{-10} \hspace{1mm} \mathrm{G} \right)^2$, 
the spectral index $n$ should satisfy 
the following relation:  
\begin{eqnarray}
n < -3 - \frac{4}{\log_{10} \left( 1 \mathrm{Mpc}/L_\mathrm{c} \right)}.
\label{eq:10} 
\end{eqnarray}

On the other hand, 
the strength of the cosmological large-scale magnetic fields is constrained 
by the CMB anisotropy measurements 
(for more detailed explanations to constraints 
on cosmological magnetic fields see Refs.\ \cite{Widrow, Kolatt}).  
Homogeneous 
magnetic fields during the time of decoupling whose scales are larger than the 
horizon at that time cause the Universe to expand at different rates in 
different directions.  Since anisotropic expansion of this type distorts 
the CMB radiation, the measurements of the CMB angular power spectrum impose 
limits on the strength of the cosmological magnetic fields.  
Barrow, Ferreira, and Silk \cite{Barrow} carried out a 
statistical analysis based on the 4-year COBE data for angular anisotropy and 
derived the following limit on the primordial magnetic fields that are 
coherent on scale larger than the present horizon: 
\begin{eqnarray} 
B(L = L_\mathrm{IR}, t_{0}) < 4.8 \times10^{-9} \hspace{1mm}\mathrm{G}, 
\label{eq:11} 
\end{eqnarray} 
where 
$L_\mathrm{IR} = \beta {H_0}^{-1}$ 
is the infrared cutoff scale of the magnetic fields.  
Here, $\beta (> 1)$ is a dimensionless constant and 
$
H_{0} = 100 h 
\hspace{1mm} \mathrm{km} \hspace{1mm} {\mathrm{s}}^{-1} \hspace{1mm} 
{\mathrm{Mpc}}^{-1} 
$ 
is the Hubble constant at 
the present time (throughout this letter we use $h=0.70$ \cite{HST}).  
Moreover, we have assumed the spatially flat Universe 
(see also \cite{Subramanian1, Tashiro1}).  
Hence, it follows from Eqs.\ (\ref{eq:8}) and (\ref{eq:11}) that 
the lower limit on $n$ is given by 
\begin{eqnarray}
n > -3 - \frac{2 \left( \log_{10} 4.8 + 3 \right)}{\log_{10} 
\left( L_\mathrm{IR} /L_\mathrm{c} \right)}.
\label{eq:12} 
\end{eqnarray}
If there exists the value of $n$ which satisfies both the 
relations (\ref{eq:10}) and (\ref{eq:12}), the value of the right-hand side 
of the relation (\ref{eq:10}) is larger than 
that of the relation (\ref{eq:12}).  In this case, using 
the relations (\ref{eq:10}) and (\ref{eq:12}), we obtain 
\begin{eqnarray} 
L_\mathrm{c} < 4.8 \beta^{-1.2} \times 10^{-5} \hspace{1mm} \mathrm{Mpc}.  
\label{eq:13} 
\end{eqnarray}
In deriving the relation (\ref{eq:13}), we have used 
$H_0^{-1} = 2997.9 h^{-1} \hspace{0mm} \mathrm{Mpc}$.  

Consequently, 
if 
the spectrum of the generated magnetic fields is slightly red 
%and the spectral index $n$ satisfies the relations 
and 
the scale $L_\mathrm{c}$ on which the field strength of the magnetic fields 
is equal to $10^{-12}$G satisfies the relation (\ref{eq:13}), 
the amplitude of the magnetic fields on 1Mpc scale 
can be larger than $10^{-10}$G at the present time 
without being inconsistent with 
the CMB anisotropy measurements.  
For example, if $\beta = 100$, from the relation (\ref{eq:13}) we can 
take $L_\mathrm{c} = 10^{-7} \mathrm{Mpc}$.  
In this case, 
it follows from the relations (\ref{eq:10}) and (\ref{eq:12}) that 
the spectral index $n$ has to satisfy the relation: $-3.58 < n < -3.57$.

%%%%%
Finally, for comparison, 
we note the case in which the spectrum of the magnetic fields 
is blue, i.e., $n > -3$.  
In the case of a blue spectrum, there exist constraints for the 
amplitude of the magnetic fields from 
the production of gravitational waves \cite{Caprini01} 
and the Big Bang Nucleosynthesis (BBN) \cite{Grasso2}.  
The former constraint is much more stringent than the latter. 
According to Ref.~\cite{Caprini01}, 
the magnetic fields generated during an inflationary phase 
(reheating temperature $T \sim 10^{15}$GeV) 
with a spectral index $n \sim 0$, the  magnetic 
fields have to be weaker than $10^{-39}$G.  
Hence the possibility of a blue spectrum is considered to be ruled out. 
%%%%%

%%%%%%%%%%%%%%%%%%%
%%%  Sec. IV
%%%%%%%%%%%%%%%%%%%
\section{Conclusion}

In the present paper 
we have considered the spectrum of large-scale magnetic fields 
generated due to the breaking of the conformal invariance of the Maxwell theory through some mechanism in inflationary cosmology.  
As a result, we have shown that 
the spectrum of the generated magnetic fields should 
not be perfectly scale-invariant but be slightly red 
so that the amplitude of large-scale magnetic fields 
can be stronger than $\sim 10^{-12}$G at the present time.

%%%%%%%%%%%%%%%%%%%%%%%%
%%%  Acknowledgements
%%%%%%%%%%%%%%%%%%%%%%%%
\section*{Acknowledgements}
The author is deeply grateful to 
Jun'ichi Yokoyama for helpful discussions.  
This work was supported in part by 
the Monbu-Kagaku Sho 21st century COE Program 
``Center for Diversity and Universality in Physics" 
and was also supported by 
a Grant-in-Aid provided by the 
Japan Society for the Promotion of Science.

%%%%%%%%%%%%%%%%%%%%%%%%%%%%%%%%%
%% thebibliography environment 
%%%%%%%%%%%%%%%%%%%%%%%%%%%%%%%%%

\end{document}